# An Owner-managed Indirect-Permission Social Authentication Method for Private Key Recovery

Wei-Hsin Chang, and Ren-Song Tsay, *Member, IEEE*

**Abstract**—In this paper, we propose a very secure and reliable owner self-managed private key recovery method. In recent years, Public Key Authentication (PKA) method has been identified as the most feasible online security solution. However, losing the private key also implies the risk of losing the ownership of the assets associated with the private key. For key protection, the commonly adopted something-you-x solutions require a new secret to protect the target secret and fall into a circular protection issue as the new secret has to be protected too. To resolve the circular protection issue and provide a truly secure and reliable solution, we propose separating the permission and possession of the private key. Then we create secret shares of the permission using the open public keys of selected trustees while having the owner possess the permission-encrypted private key. Then by applying the social authentication method, one may easily retrieve the permission to recover the private key. Our analysis shows that our proposed indirect-permission method is six orders of magnitude more secure and reliable than other known approaches.

✦

## 1 INTRODUCTION

Digital authentication has become more integrated into our daily life in this Information Age. The most widely employed authentication method is the password method, which is known to be inherently vulnerable to attacks. In a data breach investigations report [32], 81% of data breaches are hacked through stolen or weak passwords. Even the enhanced multi-factor verification method, which checks several independent pieces of evidence for authentication, may be subject to intercepting and forwarding attacks [2][3]. Therefore, researchers are anxiously looking for better alternative solutions.

The Public Key Authentication (PKA) method has been identified as the most secure authentication method available [33]. The PKA method adopts an asymmetrical cryptography system that uses pairs of keys: openly circulated public keys and owners' private keys. A private-key owner can generate a unique unforgeable signature and allow the account login server to verify whether the signature is genuine using the paired public key.

The PKA method is known to be secure if each person can safely keep the personal private key. To this issue, some have proposed a convenient and secure key management approach using Cryptographic Mobile Devices (CMDs), such as smartphones, wearables [6], or implanted chips [34]. Most CMDs adopt biometric sensors to provide a highly secure key storage and verification implementation. With the CMD solution, one can apply the highly secure PKA approach for everyday account authentications. For example, FIDO [4][56] proposes a universal password-less authentication protocol by integrating the PKA and the CMD solutions. It is becoming popular for people to use wearable CMDs to authenticate personal accounts instead of using passwords, physical keys, or cards [5][6].

However, there exists a detrimental risk. If a CMD owner loses or breaks the CMD, then the owner loses all private keys and consequently the accessibility to all corresponding accounts.

- Wei-Hsin Chang is with Deepmentor Inc., E-mail: mailggwhc@gmail.com
- Ren-Song Tsay is with the Computer Science Department, National Tsing-Hua University, Hsinchu, Taiwan. E-mail: rstsay@cs.nthu.edu.tw

Many people do lose their accounts because of losing private keys [35]. Therefore, it is critical to be able to recover private keys. In other words, we need to provide a highly secure and reliable backup authentication method for private key recovery in case of a broken or lost primary authenticator, such as CMD, discussed here.

An essential requirement for the backup authenticator is that it must be at least equally secure as the primary authenticator. Otherwise, the overall system security level is degraded since the weakest element of a system determines the actual security level [21]. Additionally, the backup authenticator needs to be very reliable. Whenever one needs to perform recovery, the backup authenticator can be reliably applicable, although there is no definite time when one may need it; otherwise, the person in need may disastrously lose all accounts.

Existing backup authentication methods can be classified into two categories, the alternative-authenticator and the original-authenticator approaches, differentiated by whether to recover the original authenticator. The alternative-authenticator approach is to register another authenticator specifically for backup authentication in advance. The original-authenticator approach is to store a backup of the original authenticator. We review both methods in the following.

Most online service providers do provide backup authentication schemes using alternative authenticators. Examples are backup code, backup account, registered phone number, or face recognition. Nevertheless, all these recovery processes are less secure than the PKA method and hence are inappropriate.

In contrast, the recently emerging decentralized cryptocurrency frameworks, such as bitcoin and ethereum, have no service providers to help to recover user accounts. Therefore, an account owner has to manage their private key to avoid permanent loss of the account access right using the original-authenticator approach. In other words, providing a secure and reliable private key backup and recovery scheme is essential to making the PKA method practical.

Since the private key is a secret created by the owner, the problem of private key backup is equivalent to the problem of protecting and recovering a secret indefinitely. Therefore, we use the term "secret" in a general sense to also cover the private key to simplify later discussion.

For security reasons, only the owner can access the secret. According to Brainard et al., there are four access-right authentication methods, i.e., something-you-have, something-you-know, something-you-are, and someone-you-know [21]. For convenience, we simply name these the *something-you-x* methods.

The first three *something-you-x* authentication methods are known to be insecure or unreliable. The something-you-have method is to store the secret at a secret place that requires something, such as a physical key or token, to open the secret place for accessing the secret. The issue of the physical key or token is its vulnerability to theft.

For theft prevention, a secret can be encrypted by a password or something-you-know. However, a simple and memorable password is vulnerable to dictionary attacks [14]. In contrast, a long, complex password is more secure but hard to memorize. Hence the long, complex password usually has to be recorded, and then the record suffers the same security issue that occurs in the method of something-you-have.

Finally, the method of something-you-are is to protect the secret using personal biometric information, which cannot be stolen or lost. However, the biometric is an exposed pattern that can attract spoofing attacks [50][60]. Also, the method lacks reliability because aging or accidents can alter personal biometrics.

The fundamental issue of the three methods above is that they all require a new secret to protect the original secret, and the new secret needs to be protected too. Therefore, they all fall into the circular protection issue and are defective.

Lately, Brainard et al. proposed a social authentication (or someone-you-know) method [21] for fallback authentication of online accounts. The social authentication method leverages the fact that the trusted contacts (trustees) of an owner of a secret can easily recognize (authenticate) the owner simply by live interaction (seeing or hearing). The advantage of the social authentication method is that the intangible social relationship cannot be stolen or lost. Since the social relationship is unique to everyone and requires no protection, the method effectively breaks the circular protection issue.

Vu et al. [52] extend the social authentication method for secret recovery by direct escrow of the secret to someone the owner trust. Furthermore, since trustees may not be trustworthy, Vu et al. combine a secret sharing method [24] to avoid betraying trustees by dividing a secret into pieces and escrowing each piece to a different trustee. Then for secret recovery, the owner simply contacts trustees and gathers a threshold number of secret shares.

Vu's method has dramatically improved all previous techniques but is still at risk of trustees' collusive attacks [52]. To eliminate the collusion attack issue, we propose a highly secure and reliable, indirect-escrow method by leveraging the non-circular protection property of the social authentication method and the effectiveness of the PKA encryption method. We briefly introduce the idea below.

We first introduce the concept of possession (where the secret is stored) and permission (the access right to the secret) to analyze the effectiveness of secret protection methods in depth.

In general, a secret protection method has to protect both the secret's possession and permission. For example, a printed secret by itself provides both permission and possession. Whoever possesses the printed secret naturally has permission to reveal the secret. If now the secret is stored in a safe box, then for any person to reveal the secret, the person will need to not only possess the safe box but also have the safe key for permission to unlock the safe box to retrieve the secret.

The three *something-you-x* secret protection methods essentially separate the permission from the possession. Nevertheless, a defect is that the permission becomes a new target to be protected and falls into the circular protection issue.

In contrast, the direct-escrow method gives each trustee partial permission and possession. Hence the method is subject to the risk of collusive attacks. The direct-escrow way is not sufficiently secure as the trustees have both permission and possession. Therefore, one idea is to give trustees only permission but not possession to fix the problem.

Therefore, we propose an improved indirect-escrow method that separately stores the possession at the owner's side but escrows the permission to trustees. Essentially, the indirect-escrow method creates encrypted secret shares using trustees' public keys and stores the secret shares locally. The owner can then recover the original secret by sending the encrypted secret shares to trustees for social authentication to decrypt the secret. The indirect-escrow method leverages the unique advantages of the PKA method and social authentication.

Most uniquely, the proposed indirect-escrow social authentication method does not need to let trustees know that they are trustees. Hence, no one possesses any secret share except the owner. In this way, the proposed indirect-escrow approach eliminates the possibility of collusive attacks and is highly reliable and secure.

However, the above-proposed approach can still be under collusive attacks after performing the secret recovery since the recovery process also exposes the list of trustees who each possess a secret share sent from the owner. After secret recovery, the situation becomes similar to the direct-escrow method and is at risk of collusive attacks.

To further eliminate this collusion possibility, we devise an improved indirect-permission method that pre-encrypts the secret to be protected using a randomly selected symmetric key and perform an indirect-escrow of the random key instead of the secret itself. Therefore, trustees know only the tentative random symmetric key even after secret recovery, not the secret itself. In this way, the collusive attack is avoided.

The most significant contribution of our proposed indirect-permission method is that it is straightforward to implement and is highly secure against all known attacks, such as theft, password cracking, biometric spoofing, and collusion. Details are to be elaborated on in later sections.

This paper is organized as the following. We first review related work in Sec. 2 and present our proposed methods in Sec. 3. Then, we summarize in Sec. 4 our experimental results and compare them quantitatively with other approaches in Sec. 5. Finally, we conclude the paper in Sec. 6.

## 2 RELATED WORK

As discussed, existing backup authentication methods can be classified into two general approaches, the alternative-authenticator, and the original-authenticator approaches, depending on whether an owner has an alternative path to access the account. We now review related work in the following and compare the associated security properties.

## 2.1 The Alternative-Authenticator Approach

An alternative-authenticator approach is to register another secret in advance for backup authentication. This approach is common in password-based authentication systems. However, applying this approach to the PKA system will degrade the security gained by the PKA because the alternative secrets used are generally weaker. We review representative methods in the following.

### 2.1.1 The security-question method

The security-question method probably is the first commonly adopted account recovery scheme. To recover an account access right, the service provider asks the requester for certain pre-registered personal information, such as the first telephone number used, the father's middle name, etc. Although answers to these questions are personal and supposedly unforgettable, this scheme is deemed insecure as the answers are generally easy to guess due to the small answer space. Also, one may easily find the answers on the owner's social media records. Furthermore, asking personal questions intrudes into the user's privacy [31]. Also, since the same answers and questions may be used on different sites, this method is intrinsically defective [36].

### 2.1.2 The out-of-band method

Besides personal knowledge, out-of-band communication accounts such as personal e-mail or phone accounts can also be used as backup authenticators. The out-of-band method provides higher security than the security-question method because of the requirement for an additional password or proof of physical ownership. For this method, usually, an account owner pre-registers an e-mail address or phone number. If the account password is forgotten, the owner may ask the service provider to send a password reset link or verification code to the pre-registered personal e-mail account or the personal phone using SMS (short message service).

This method assumes that the e-mail or SMS communication method is secure. However, in reality, an attacker may hack e-mail servers [36] or eavesdrop on the unencrypted channel to retrieve e-mails. Similarly, one may intercept the verification code sent by SMS [38]. Cases of implanting malware on a client device to recover the owner's E-mail/SMS [31] and the Man-in-the-Middle attack (MitM) to intercept the recovery code [3][37] have been reported. Although the PKA method is as secure as the original authenticator, using this out-of-band method for backup recovery will significantly lower the system's security level.

### 2.1.3 The backup code method

For this method, the service provider issues a backup code and has the account owner keep the code secret [57]. The owner usually prints out the code to prevent malware attacks. When recovering an account, the owner submits the code for authentication. The printed code is at risk of theft, and MitM (Man-in-the-Middle) attackers may fish the code presented. The security of the approach depends on how well the owner protects the code.

### 2.1.4 The face recognition method

Some service providers adopt the face recognition method for account login [53][54] and account recovery. For this approach, the service providers store each owner's biometric information online for identification. One prominent issue is that the centrally managed approach is infamous for invading the user's privacy. Another serious challenge is that a person's face may vary due to aging or accident, and the changes may cause a person to fail to access a personal account [49]. Additionally, Xu presented a face spoofing method by constructing a 3D face model using the target's online photo and demonstrated that the face recognition method is vulnerable to spoofing attacks [60].

### 2.1.5 The social authentication method

Brainard et al. [21] proposed a social authentication method to help users recover their accounts when losing one of two factors in a two-factor authentication system. Unlike the other recovery methods that require the account owner to provide proof of identity, which is vulnerable to being stolen, as discussed before, the owner's identity can be confirmed by the one he trusts (trustee) simply by a phone call.

In case of losing the account token, the social authentication method has the account owner request the service provider issue a unique recovery code to a pre-registered trustee. The account owner then calls the trustee by phone. The trustee releases the recovery code to the owner if the caller is confirmed to be the owner. The owner then forwards the recovery code and a second-factor code to the service provider to recover the account. This trustee-owner phone-call authentication process is more challenging for hackers to steal or spoof than other methods.

However, since a second-factor code is required, this method does not apply to most systems that are single-factor systems. In this case, the trustee may easily impersonate the owner and access the account using the received recovery code to replace the single authentication factor.

To apply to the password-based single-factor systems, Schechter et al. proposed a multi-trustee social authentication method deployed on Windows Live ID [23]. The method avoids relying on a single trustee by requiring multiple special recovery codes for account recovery. Unfortunately, this method has a loophole as the service providers, for usability reasons, typically provide the list of trustees to the account owner as a reminder for whom to contact for recovery. The issue is that whoever knows the list may organize the trustees for collusive attacks. Additionally, real-world experiments showed that 45% of the trustees could be cheated by someone close to the account owner, even having received specific verification instructions from the service providers [23]. This fact implies that phone calls or face-to-face meetings are not reliable for validation.

Facebook has proposed another social-authentication-based account recovery method called trusted contact [22], adding some defense mechanisms against unreliable trustees. First, the system requires the person who requests account recovery types in a trustee's name to show the complete trustees list. This requirement reduces the probability of revealing the list of trustees to attackers. However, an attacker may try the name of someone close to the target owner and break into the account [39]. Therefore, the system also sets a 24-hour waiting period before granting the account access right to the one that gathers enough recovery codes. If someone logs in to the account with the old password during this period, the account recovery is deemed fake and denied. However, this defense mechanism highly depends on how often an owner logs in to the account. Furthermore, in case the owner does forget the password and cannot log into the account, then, obviously, the collusion of trustees will prevail.

## 2.2 The Original-Authenticator Approach

This approach assumes that only the original authenticator can be used to access the account. The PKA approach is typical in this category. Therefore, the focus is to provide a secure backup and recovery method for the original authenticator. Depending on who possesses the backup, we classify the backup methods into the owner-possessed and the direct-escrow approaches. We then further classify each category into sub-categories according to who owns the permission.

### 2.2.1 The owner-possessed approach

a) Local storage (something-you-have): The most intuitive backup approach is to have a physical copy and store the document in a local offline place. This approach requires no password to remember, and the offline backup is resistant to online attacks. The backup can be on paper [15] or metal (good for long-term backup) [40]. However, whoever possesses the printed private key naturally has permission to reveal the private key and access the account. Therefore, this method is vulnerable to theft.

For theft prevention, the backup must be protected by authorized permission. An example is a key or password to a safe box. However, the new permission needs to be protected, and the additional protection becomes a circular issue. Another risk is that the locally stored backup can be destroyed permanently in a fire or disaster.

b) Password protection (something-you-know): Password-protected backup is resistant to theft. For example, a digital wallet containing private keys is protected by a password [16], and a backup is a copy of the wallet stored in another place. The security of this method highly depends on the strength of the password. Although the password space theoretically is huge, researchers found that specific patterns exist in human-made passwords [42]. Therefore, an attacker with the backup may guess a password by applying the known patterns. Recent research showed that one could guess in 16 days 40% of real-world passwords simply by using a recently leaked password set with an NVidia GeForce GTX 980 Ti [42].

Although one may generate a long and randomized password to defend against guessing, a complex password is hard to memorize and must be recorded on something and stored somewhere. The additional record again becomes a circular protection issue.

Some generate private keys via passwords [17][47] and allow the owner to recover the lost private key using the corresponding password. However, since the password approach is fundamentally not secure, the security level is downgraded. For example, by password scanning, M. Vasek et al. [48] identified 884 active bitcoin accounts worth around $100K in 2015.

c) Biometric protection (something-you-are): Instead of password protection, the backup can be protected using biometric methods. It is harder to reproduce a person's biometric than a password. In practice, an owner may secure private keys in a device equipped with a biometric authenticator [52]. However, if the entrusted device breaks, as everything has a limited lifetime, the owner loses all the private keys and access rights to associated accounts.

Alternatively, some proposed biometric-generated private key methods [50] are resistant to guessing because of the complexity of the biometric information. However, attackers may spoof the owner's biometric data and generate the desired private key [50][60].

Another general concern of the biometric approach is that no one can control the long-term biometric variations caused by aging, disease, or accident [49].

### 2.2.2 The direct-escrow approach

a) Escrow to the server(s): Instead of keeping private keys locally, as in the owner-managed approach, private keys can be stored online and accessed using a password or a 2-factor authentication method [43]. However, all the password or multi-factor-related issues, such as the MitM attack issue, occur here [44]. Additionally, since the online servers store many private keys, they become the target of attacks [46].

To overcome the server attack problem, S. Jarecki et al. [45] proposed a "Password Protected Secret Sharing (PPSS)" approach, which divides each private key into pieces by secret sharing [24] and stores them in different servers. The owner may recover the private key after valid authentications over a threshold number of servers. In other words, an attacker must hack over the threshold number of servers to steal the private key; hence, the multiple-hacking task is much more difficult.

For further security enhancement, some proposed a proactive secret sharing method [58] that periodically renews the secret sharing process to defend against server hacking. This method is based on the observation that attackers cannot break enough servers in a short time and hence proposes a periodic renewal process to invalidate the attacker-gained information. Fundamentally, how to remember and protect the server login method, such as password, is an unresolved circular issue.

b) Direct-Escrow to trustees: L. H. Vu et al. proposed a social-authentication-based private key backup method for the distributed online social network (DOSN) [51]. The private key is divided into secret shares by secret sharing [24], and each share is stored in the selected trustee's server, which is assumed to be well protected by the trustee. To recover the private key, the owner contacts the trustees and retrieves the private key after collecting a threshold number of secret shares. The essence of this social authentication approach is that it does not require new permission to be protected and hence eliminates the circular protection issue. The reason is that social relationship is a naturally decentralized authentication scheme that is very difficult to reproduce. The only possible threat to this approach is that the trustees may collude to steal the owner's private keys.

Some attempted to design a better trustee selection algorithm to resolve the collusion issue. Vu [51] suggested selecting trustees from different sets of friends and encrypting the secret shares using passwords. The problem is that the trustees may eventually know each other and collude with others to recover the key. Also, the encryption password suffers the same issues as the password protection methods discussed before.

Nojoumian et al. [59] proposed a reputation-based secret shares distribution scheme to identify reputable secret holders iteratively. An issue with this approach is that the private key may leak through collusion before truly reputable trustees are identified.

In summary, we observe that the critical issue for potential collusive attacks is because trustees directly possess the secret shares and have the chance for collusion. To be free from collusive attacks while avoiding circular permission issues, we propose improved methods to be elaborated below, with no trustees directly possessing secret shares.

## 3 THE PROPOSED SECRET BACKUP APPROACHES

Before we elaborate on our proposed backup method, we first introduce the involved factors and the models of security and reliability. Then after describing our proposed approach, we quantitatively analyze our method's security and reliability measures and compare our results with other approaches.

### 3.1 Notations

In this section, we first identify involved parties and channels and examine the secret protection process. We first introduce a few terms to simplify the discussion of the encryption process. We also formally define the objective of security and reliability.

*3.1.1 Parties*

We call the owner of the private key simply the *owner*, who possesses the private key (or the secret key, *SK*) and generates the backup.

To use social authentication, the owner selects some *trustees* from his contacts, and the remaining contacts are simply called *contacts*. The chosen trustees are supposed to be able to identify the owner via interactions and help to recover the owner's private key.

*An adversary* is the one who attempts to steal the owner's private key and break into his account. Generally, the adversary has a low probability of physically stealing the owner's backup. In contrast, traditional backup methods are vulnerable to being stolen, as we have discussed before. With the computing power of current technologies, we assume that the adversary cannot break the owner's public key encrypted information, and the adversary cannot fully emulate the interaction style and details of the owner. The potential threats of quantum computing shall be investigated in future work.

*3.1.2 Communication Channels*

**User-Server**: We assume the channel between a user to a server is secured by the PKA method. In other words, we assume no adversary can eavesdrop or forge messages over such a channel.
**Peer-Peer**: We assume that every owner can establish a secure communication channel with trustees. No adversary can eavesdrop, forge or intercept messages in the transactions.

*3.1.3 Encryption Notation*

We use the notation "$\odot$" to denote the operation of encryption. For example, "PK$\odot$information" gives the result after PK encrypts the information.

*3.1.4 Evaluation Measures*

Here, we first define the security and reliability measures and later use them to evaluate the quality of backup approaches. The **security** measure is the probability that adversaries fail to retrieve the backup, and the **reliability measure** is the probability that an owner can successfully recover the backup. A backup fails if it is either not secure or not reliable.

### 3.2 Assumptions

We assume that PKA will be the primary authentication method in the future because of its security and mature key management solution. The owner can openly access trustees' public keys through online identity servers without the need to inform trustees. Additionally, the access is not constrained to specific servers; hence, no service providers can control or dictate the operation of our proposed method.

Since we assume the adversary could not break into the CMD, an owner needs to perform recovery only when his CMD is lost or not functional. Generally, the owner and trustees tend to protect their private keys well and rarely change them.

Finally, we assume the owner and trustees have a secure computing environment for processing the encryption and decryption.

### 3.3 The Direct-Escrow Method

We first demonstrate the weakness of the traditional direct-escrow approaches. As discussed in section 2.2.2, the direct-escrow method is a practical approach to avoiding the circular protection issue by using the social authentication technique. Specifically, the owner generates secret shares, $\overline{sk}$, of the to-be-protected private key *SK* using the secret sharing method [24] with parameter (k, n), where n is the number of trustees and k is the recovery threshold. We define the set of secret shares, $\overline{sk} = \{\overline{sk_i}, i = 1 \sim n\}$. Each share is then stored at a trustee's server; each trustee has partial possession and permission of *SK*. The owner only needs to gather secret shares from at least *k* trustees to recover the original private key *SK*.

The main weakness of the approach is that, since each trustee knows they are a trustee of the owner, the trustees may attempt to contact each other and collude to steal *SK*. Furthermore, a research report [23] revealed a non-negligible possibility for the trustees to be cheated into submitting secret shares. For this approach, an adversary only needs to cheat or collude *k* trustees to steal the owner's *SK*. Although the direct-escrow method has excellent improvement over the traditional methods, the approach is still vulnerable to collusive attacks, and its security can be further improved.

### 3.4 Our Proposed Algorithms

In this section, we introduce our proposed indirect-escrow and indirect-permission backup methods, which mainly reduce the collusion possibility and are both highly secure and highly reliable.

*3.4.1 Indirect-escrow Method*

First, our indirect-escrow method separates the management of permission and possession and assigns only the permission of the backup to trustees. The idea is to encrypt the secret shares $\overline{sk}$ using the public keys of the selected trustees $pk = \{pk_i, i = 1 \sim n\}$, i.e., perform $pk \odot \overline{sk} = \{pk_i \odot \overline{sk_i}, i = 1 \sim n\}$. This method leverages the fact that each person's public key is publicly accessible for backup preparation with no notification required.

Note that the owner, not the trustees, keeps the encrypted secret shares $pk \odot \overline{sk}$ and submit them to the trustees for decryption only when needing recovery of *SK*. Since trustees do not possess the secret, we call this approach the "indirect-escrow" method.

This indirect-escrow method does prevent trustees from direct collusion as they do not have secret shares. However, since the owner cannot obtain the private keys of the trustees for decryption, they need to send the secret shares to the trustees when needing recovery. At that time, the trustees will possess the secret shares after decryption, and then the situation becomes similar to that of the direct escrow as the trustees may collude to get the whole secret. Therefore, the indirect-escrow approach still has a security hole, although it is much more secure than the direct-escrow method.

*3.4.2 Indirect-permission Method*

For further improvement, we propose an indirect-permission method that prevents the trustees from possessing the secret *SK* even after the owner requests recovery. The main idea is to pre-encrypt the to-be-protected secret *SK* using a randomly selected symmetric key *RK* and perform an indirect escrow of the random symmetric key *RK* instead of the secret. Since *RK*, which is indirectly escrowed, is the permission to the owner-possessed encrypted secret *RK*⊙*SK*, we, therefore, name this approach the indirect-permission method.

Even with collusion, the trustees can obtain only RK but not the encrypted secret *RK*⊙*SK during the recovery phase*. After recovery, the owner should have the *SK* and can update a new random key and a new set of trustees for a new backup setup. At the same time, the owner deletes the previously encrypted secret. Therefore, the indirect-permission method solves both the circular permission issue and the collusion problem.

For our approach, the trustees do not know that they are selected trustees until receiving recovery requests, while for the direct-escrow approach, trustees know who is the owner of the escrowed backup. Since, for our approach, anyone may request recovery, the trustee must validate the actual ownership of the request to avoid the attacker's false request. Therefore, in our approach, we further require the owner digitally sign each secret share along with a decryption instruction to the trustee using the owner's private key so the trustees can use the owner's public key to verify the signature and know how to proceed.

In the following, we elaborate on the backup and recovery algorithms of the indirect-permission approach. More implementation details are discussed later.

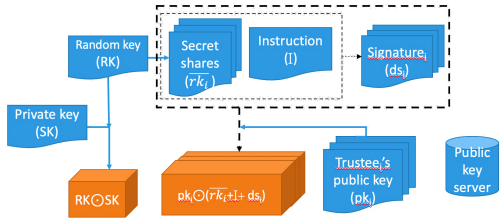

Fig. 1: The architecture of the proposed indirect-permission backup method.

**The Backup Phase:**
The architecture of the backup algorithm is illustrated in Fig. 1, and the details are described below.

*(1) Encrypt the target secret*
The owner generates a random symmetric key, *RK*, and encrypts *SK* into *RK*⊙*SK*, as shown in Fig. 1. The specification and length of *RK* conform to contemporary standards.

*(2) Select trustees and perform secret sharing*
The owner selects *n* of their contacts as trustees and sets a recovery threshold *k* (*k* ≤ *n*) for fault tolerance. Let trustees form a set T = {trustee$_i$, i=1~n}. The owner performs (k, n) secret sharing on *RK* using Shamir's secret sharing method [24]. Suppose that these secret shares are $\overline{rk} = \{\overline{rk_\iota}, i = 1 \sim n\}$. Then *RK* can be recovered after combining any *k* secret shares in $\overline{rk}$.

There are a few known trustee selection strategies [51][59], all of which can work well with our approach. In practice, we suggest that the owner evaluates the security and reliability level of trustee candidates under different conditions and then chooses the ones with higher security and reliability.

*(3) Recovery instruction to trustees*
To smooth out the recovery phase, besides the encrypted backup permission $\overline{rk}$ we also attach a recovery instruction *I* to guide trustees on what to do when receiving the recovery request. For instance, the recovery instruction may contain the owner's name for verification and the required communication format. For example, the owner may ask the trustees to verify the owner in person or allow verification through text or video messaging. Furthermore, the owner may assign legal agents in the recovery instruction if the owner becomes communication disabled.

Although we cannot expect everyone to follow the instruction, we believe most of them will follow the instruction for recovery since trustees are those the owner trusts.

*(4) Digital signature*
A critical step during the recovery phase is to have trustees identify whether the submitted secret shares for recovery genuinely belong to the one who requests recovery. For this purpose, we have the owner use their private key *SK*, encrypt the hashed result of each combination of *rk$_i$* and *I*, and generate a digital signature package, $ds_i = SK \odot \text{hash}(\overline{rk_\iota}, I)$.
In the recovery phase, this *ds$_i$* will help *trustee$_i$* validate the ownership of the secret share $\overline{rk_\iota}$ and the instruction *I*.

*(5) Download trustees' public keys*
Download each trustee's public key *pk$_i$* from her account or interaction history.

*(6) Encrypt using trustees' public key*
We then use *pk$_i$* to encrypt each set of $\overline{rk_\iota}$, *I*, and *ds$_i$*. We define $\boldsymbol{pk} \odot \overline{\boldsymbol{ss}} = \{pk_i \odot (\overline{rk_\iota}, I, ds_i), i = 1 \sim n\}$. After encryption, the owner stores *RK*⊙*SK* and $\boldsymbol{pk} \odot \overline{\boldsymbol{ss}}$ for backup and deletes other unnecessary information, such as *RK*. Since the temporary key *RK* is destroyed, the owner may safely print out the backup *RK*⊙*SK* and $\boldsymbol{pk} \odot \overline{\boldsymbol{ss}}$ into QR codes or other forms or even make several copies without fear of being decrypted.

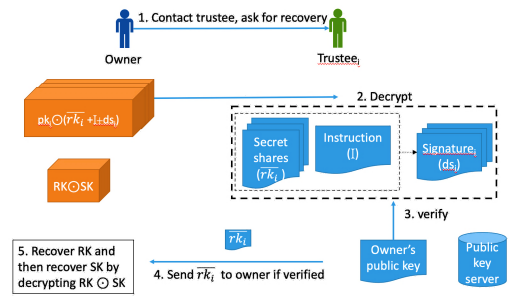

Fig. 2: The architecture of the proposed key recovery algorithm.

**The Recovery Phase:**
The architecture of the recovery algorithm is shown in Fig. 2, and the details are described below.

*(1) Request for recovery*
For key recovery, the owner sends the encrypted secret share with instrcution $pk_i \odot (\overline{rk_\iota}, I, ds_i)$ to trustee$_i$ via a secure channel. Note that the backup does not contain the trustee's name for security reasons. Practically, the owner may send the whole set $\boldsymbol{pk} \odot \overline{\boldsymbol{ss}}$ to trustee$_i$ for recovery and let trustee$_i$ attempt decrypting all files and find the specific $pk_i \odot (\overline{rk_\iota}, I, ds_i)$.

### (2) Trustee follows the recovery instruction

After successful decryption of $pk_i \odot (\overline{rk_\iota}, I, ds_i)$, trustee$_i$ reads and follows the recovery instruction $I$, which we assume the trustee is willing to follow.

### (3) Verify ownership

Trustee$_i$ downloads the owner's public key, $pk_{owner}$, from the owner's account, whose address can also be found in instruction $I$. The trustee$_i$ applies $pk_{owner}$ to decrypt $ds_i$ into some value, say $v$, which shall equal the hash value of $(\overline{rk_\iota}, I)$ as expected. If the equality is validated, then the secret share $(\overline{rk_\iota}, I, ds_i)$ is verified to be authentic by the account owner. Then the trustee checks whether the requester who contacts him is the account's owner. Note that trustees are supposed to recognize whether the requester is the owner through interactions with the requester.

### (4) Decrypt secret shares

If the ownership is verified ok, the trustee sends the decrypted secret shares $\overline{rk_\iota}$ back to the owner for recovery of the temporary key.

### (5) Recover the owner's private key

The owner repeats steps 1~4 until collecting k secret shares. With the accumulated secret shares, the owner reconstructs the temporary key $RK$ following Shamir's secret sharing method and then applies the reconstructed $RK$ to decrypt the backup copy of $RK \odot SK$ and recover the private key $SK$.

Since the trustees, after recovery, are now exposed, they may collude to regenerate the temporary key $RK$. Nevertheless, they will need to steal the encrypted backup $RK \odot SK$ to recover the private key $SK$. To eliminate such a possibility, the owner shall destroy all backups after each recovery and choose a new temporary key with a new set of trustees for a new run. Therefore, the proposed indirect-permission approach can be highly secure and reliable.

## 4 SECURITY AND RELIABILITY ANALYSIS

Both security and reliability are critical measures of any successful secret backup and recovery approach. However, most works consider only security measures. In reality, either the security or the reliability fails, and the recovery fails.

For reliability, we are talking about the probability that an owner can complete the process and truly recover the owned private key when needed. In contrast, security is about the possibility an attacker fails to obtain the owner's private key. Both security and reliability are essential for any recovery approach. In the following, we first separately analyze the security and reliability. Finally, we combine them into a new integrated failure measure and then suggest how to choose parameters for the proposed approach optimally.

### 4.1 Security Analysis

In this section, we analyze all possible attack scenarios for our approach. First, an adversary must physically steal the owner's backup files $RK \odot SK$ and encrypted secret shares $pk \odot \overline{ss}$. We assume the adversary has a probability $P_{steal}$ to steal the backup successfully. Since the temporary key is destroyed after the backup process, the adversary cannot decrypt the encrypted backup $RK \odot SK$ and the secret shares $pk \odot \overline{ss}$ even if they physically get the backup files. Therefore, the adversary must locate enough trustees to decrypt the owner's private key $SK$.

To decrypt the stolen $RK \odot SK$, the adversary must collect at least $k$ random key secret shares $\overline{rk_\iota}$'s. However, only the designated trustee has the private key to decrypt $pk_i \odot (\overline{rk_\iota}, I, ds_i)$. Therefore, the only way to decrypt the backup is to find the trustees and cheat enough trustees successfully or convince enough trustees for collusion.

An adversary may guess the owner's contact list from the owner's social network profile if it is public. If the profile is private, Jin et al. [62] proposed a "mutual friend attack" method that utilizes the mutual friends of the owner. Jin's experiments showed that an adversary could simply locate 60~95% of a target's contacts using the mutual friend attack method. However, in reality, for our approach, a trustee may not even appear on the owner's online social network. At any rate, we assume that an adversary can identify all owner's contacts in the worst case.

When an adversary requests the owner's contact for decryption, the person being contacted may have the following reactions. Typically, an owner's connection may simply ignore the request (assume the probability is $p_1$) or notify the owner of the suspicion (assume the probability is $p_2$). Similarly, when a trustee is being contacted, they may simply ignore the request (probability $q_1$), notify the owner (probability $q_2$), or be fooled (probability $q_3$).

Generally, the attack will fail if anyone contacted by the adversary feels suspicious and notifies the owner. The owner can then change the random key and secret shares to break the attack since the stolen backup file is now invalid. Practically, it is implausible for an adversary to request illegal decryption without causing any suspicion.

Then we have the following worst-case probability $P$ for a successful adversary. The details of derivation are listed in the appendix.

$$P = P_{steal} \times \frac{1}{\Omega} \sum_{i=k}^{n} \sum_{j=0}^{N-n} C_j^{i-1+j} \times C_{N-n-j}^{N-i-j} \times p_1^j \times C_{i-k}^{i-1} \times q_3^k \times q_1^{i-k} \tag{1}$$

where,
$\Omega = C_n^N$,
$N$ = the total number of contacts,
$C_j^i = \frac{i!}{(i-j)!j!}$, a binominal coefficient.

Note that $n$ is the number of trustees and $k$ is the secret share recovery threshold.

The above equation is complicated and can be approximated as

$$P = P_{steal} \times \left(\frac{n}{N} q_3\right)^k, \tag{2}$$

where $q_3$ is the probability of a trustee being fooled. The equation makes sense as it implies that an adversary succeeds if they successfully steal the secret shares ($P_{steal}$) and choose just the right number ($k$) of correct trustees ($n/N$), and every trustee is easily fooled ($q_3$).

Generally, the numbers $p_i$ and $q_i$ are not controllable and can only be improved through education. Therefore, to achieve the desired security level, we should adequately manage the choice of $n$ and $k$. Since $N \gg n$ and $\left(\frac{n}{N} q_3\right) \ll 1$, with a proper choice of $n$ and $k$, we may easily achieve the target security level.

### 4.2 Reliability Analysis

Since the encrypted backup $RK \odot SK$ is stored on the owner's side, it is critical not to lose the backup. Fortunately, unlike the local storage method that needs a secret place to store the backup, our backup is very secure. Even if an adversary successfully steals the backup, he must first fool enough trustees to recover the temporary key to unlock the secret. The probability is very low, as we have just discussed. Therefore, the owner can be more flexible in selecting the storage forms and locations. An owner can make duplicated copies of the encrypted backup and store them in various places. Thus, it is implausible to lose all backups simultaneously unless large-scale disasters occur.

Another threat to recovery failure is the owner forgetting who the trustees are. Since only the owner knows who the trustees are, without knowing enough trustees, the recovery will fail. However, in this case, the owner may simply ask all candidates from the contact list and slowly identify trustees for recovery. Therefore, failure in this step is unlikely.

The reliability is lowered if trustees are unavailable in the recovery stage. Although the relationship between the owner and trustees cannot be stolen or forgotten, the trustees can be dead or lost at contact for recovery. Nevertheless, a unique advantage of our approach is that the owner can renew the backup process and elect new trustees any time she wishes since the selection process does not require consent from any trustees.

There is a possibility that a trustee may refuse to cooperate and decrypt the backup shared secret even being asked by the authorized owner. Practically, if a trustee becomes untrustworthy, the owner can simply replace the trustee with a new candidate without the need to inform the new and old trustees.

In case trustees change their private keys, although this rarely occurs because of the enormous effort to register a new key to all accounts, we propose to provide a monitoring program to inform the owner if someone changes public keys. To be safe, each owner should renew backups whenever receiving the key change notification.

Besides proactively renewing backups, an owner can enhance reliability with the parameters of secret sharing $(k, n)$. With the choice of $(k, n)$, an owner can tolerate at most $(n-k)$ unavailable or unwilling trustees. We use the same analysis methodology in security estimation by referencing general statistics and experimental data.

Suppose that the probability of a trustee who is unavailable or unwilling is $U$, then the recovery failure rate, $Q$, is

$$Q = \sum_{i=n-k+1}^{n} C_i^n \times (1-U)^{n-i} \times U^i \qquad (3)$$

As shown in equation (2) and equation (3), we find that the secret sharing parameters $(k, n)$ essentially determines the quality of our approach's security and reliability. There is a tradeoff between security and reliability. Generally, a higher $k$ value leads to a more secure but less reliable result. With the same $k$ value, the larger $n$ value leads to a higher attack success rate. Therefore, it is imperative to balance security and reliability by properly selecting $(k, n)$.

### 4.3 Recovery Failure Rate Analysis

As discussed above, a successful recovery approach should be secure and reliable. For a more precise evaluation of any recovery approach, we define the recovery failure rate $F$ as the probability that an approach is either not secure or not reliable.

$$F = 1 - (1-P) \times (1-Q) = P + Q - PQ, \qquad (4)$$

where
$P$ = the attack success rate in equation (1)
$Q$ = the recovery failure rate in equation (2)

Since, in general, $P, Q \ll 1$ and $PQ \ll P, Q$, we have
$$F = P + Q \qquad (5)$$

### 4.4 Real World Parameters

We adopt appropriate parameters to estimate the failure rate $F$ by referring to available statistical or published research data. We assume a reasonable range of values for analysis for those with no referenced data. The real-world data references are more realistic, although sometimes they differ from theoretical numbers. For example, a long enough password is theoretically hard to guess, but in reality, people tend to use some common words for convenience. Therefore, a test on real-world password sets is closer to the realistic security level of the password approach [42].

In equation (1), there are parameters $P_{steal}$, $p_1$, $q_1$, $q_3$, and $N$ to be defined. For the value $N$, we simply adopt the average number of friends on Facebook, i.e., 404 [63].

For the probability of successfully stealing backup, $P_{steal}$, we adopt the reported burglary rate. However, private key backups are generally more seriously protected, and very few thieves aim to steal encrypted secrets. As a reference, the recent FBI statistics show that in the USA, the burglary rate is 274 cases per 100,000 inhabitants [61] or 0.274%, which is used as the worst case of the backup stolen rate $P_{steal}$.

For the reaction of contacts, we refer to a real-world social authentication experiment from Schechter et al. [23]. The report shows that as high as 45% of trustees can be fooled if the adversary knows the owner well and can get hints from the owner on how to cheat trustees. Although in our case, no owner will work with an adversary to fool trustees, we still take 45% as the worst-case value for $q_3$. For $q_1$ and $q_2$, we assume each is equal to $(1-q_3)/2$. Similarly, we assume that the probability $p_1$ for a regular contact to ignore the adversary's request is similar to a contact being fooled by the adversary, i.e., we assume $p_1=q_3$.

Then in equation 2, only the trustee unavailable rate U is to be identified. Although the actual $U$ value is known only until our method's real deployment, we assume 0.1% failed trustees for later calculations.

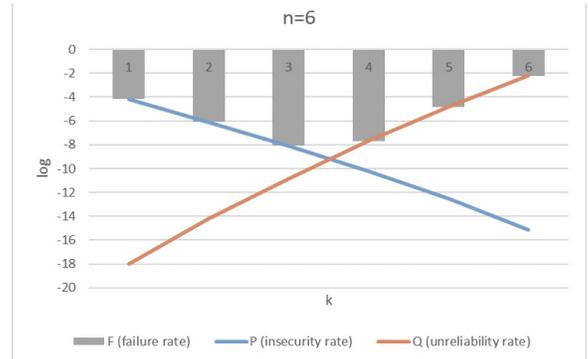

Fig. 3: The failure rates of different recovery thresholds for $n = 6$. The lowest failure rate occurs when $k = 3$.

### 4.5 Failure Rate Optimization of (k,n)

We then compute the probabilities $P$ (insecurity rate), $Q$ (unreliability rate), and $F$ (failure rate) using the values quoted in the previous subsection. For example, we show in Fig. 3 the failure rates of different recovery thresholds $k$'s for the choice of 6 trustees.

As shown in Fig. 3, there is a tradeoff between $P$ and $Q$. Note that the minimum failure rate is $10^{-8}$ and occurs when $k$ equals 3. Therefore, to be the most secure and reliable, the owner should set the recovery threshold to 3 if he has six trustees.

Therefore, for each choice of $n$, one may calculate the optimum $k$. Hence, we show in Fig. 4 the minimum failure rate and the corresponding $k$ value for each option of the number of trustees $n$. Practically, one may determine the target allowable failure rate and use this table to choose the optimal parameters of $(k, n)$.

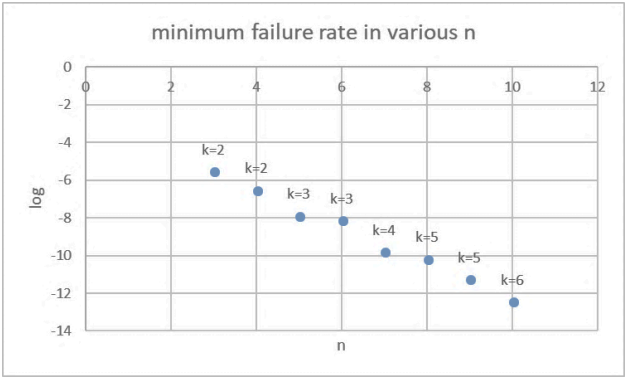

Fig. 4: The minimum failure rate for each different choice of the number of trustees $n$.

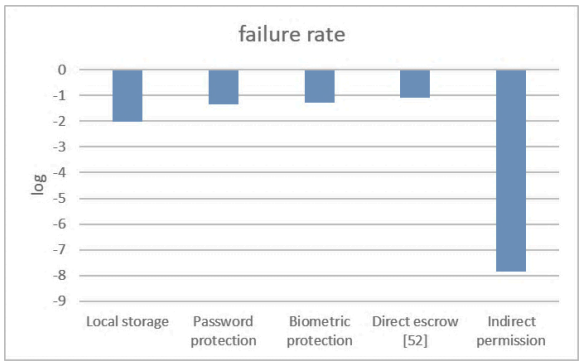

Figure 5: The failure rate comparison.

## 5. COMPARISON

In this section, we compare our proposed indirect-permission method with other private key backup and recovery methods using the recovery failure rate measure in equation (4) based on the same estimated $P$ (insecurity rate) and $Q$ (unreliability rate) measures.

First, we take the paper wallet local storage solution [6] as an example. The paper wallet method prints the private key as a QR code and stores the code offline. For this approach, once the backup is stolen or lost, then the backup is considered a failure since anyone with the QR code can retrieve the private key. If we use the same burglary rate, 0.274%, to represent the worst-case stolen probability as discussed in section 4.4, the computed failure rate is 0.55%, much higher than our approach.

As for the password protection method, an offline backup is encrypted by a password [16]. An adversary must steal the backup and perform a brute force attack to guess the password. Recent research reported that one could guess in 16 days 40% of real-world passwords based on a recently leaked password set using an NVidia GeForce GTX 980 Ti [42]. Therefore, the attack success rate is equal to the stolen rate multiplies the password guessing rate, i.e., 0.4*0.00274=0.0011. On the other hand, the loss rate can also be estimated by the probability that an owner forgets the password. A large-scale survey of password-using habits indicates that 4.28% of users forgot their passwords over three months [64]. Therefore, the recovery failure rate is at least 4.28%, a very high number.

For biometric protection, an offline backup is stored in a biometric-enabled device, such as a fingerprint USB [51]. We take the fingerprint solution as an example for comparison since both the spoofing and the anti-spoofing techniques are relatively mature. In this case, an adversary may steal the backup and spoof through the owner's fingerprint. Therefore, the attack success rate is equal to the backup stolen rate multiplies the spoof success rate, about 5%, based on an estimate from a fingerprint liveness detection competition [65]. In this case, the loss rate is the anti-spoofing algorithm's false rejection rate, which is about 5%, as shown in [65]. Therefore, the total recovery failure rate is 5.02%, a very high number too.

For the direct-escrow method, we refer to Vu's work [52], which encrypts the secret shares using passwords and escrows them to trustees. Since the adversary knows all trustees, a successful attack can be achieved by cheating or colluding enough trustees without causing suspicion and by successfully guessing the passwords. The probability of successful cheating or collusion without suspicion essentially can be computed by equation (2) but with $N$ replaced by n. The loss rate can be calculated by equation (3) with the same $F$. For fair comparisons, we set the same total number of trustees and the recovery threshold as ours. Then we have an 8.31% failure rate, which is amazingly high due to the easy collusion possibility.

Finally, we list the recovery failure rate of each approach for comparison in Fig. 5 and Table 1 based on (3,5) secret sharing. Note that Facebook also adopts (3,5) secret sharing for its trusted contact framework [22], and hence the (3,5) parameters shall be a good reference for comparison.

In summary, the failure rate of our proposed indirect-permission approach is about six orders of magnitude better than all other results. This fact indicates that our method is both highly secure and reliable.

It is worth noting that the failure rate of both the password and biometric protection are higher than the local storage approach because both approaches emphasize security but not reliability. Therefore, the password and biometric protection approaches are more appropriate for the primary authenticator. On the other hand, the direct-escrow [52] emphasizes reliability more, while its lower security measure severely degrades the failure rate.

TABLE 1
Failure rate comparison

| Approach | Failure rate |
|---|---|
| **Local storage** | 0.55% |
| **Password protection** | 4.39% |
| **Biometric protection** | 5.02% |
| **Direct-escrow** | 8.31% |
| **Indirect-permission** | 0.0000014% |

## 6. CONCLUSION

This paper proposes indirect-escrow and indirect-permission methods for private key backup and recovery. Unlike previous approaches that keep both the possession and permission of the private key backup to either the owner or trustees, our approach lets trustees have only the permission of the backup while the owner has the possession. Our approach is highly secure due to the difficulty of locating all trustees and is highly reliable because of the redundancy of trustees. We also propose a backup failure rate measure that considers both security and reliability and suggests the optimal choice of the threshold number for each choice of the number of trustees. According to the failure rate analysis, our approach is about six orders of magnitude better than other best-known results with the same number of trustees. Therefore, we conclude that our approach provides a very secure and reliable private key backup and recovery method, which is essential for the public-key-based authentication infrastructure. The method can also be applied to protect secrets other than private keys. The indirect-permission approach can also be extended to be a multi-layer indirect-permission approach or adopt multiple sets of trustees for even better security protection.

## ACKNOWLEDGMENT

This work was supported in part by Taiwan NSTC grant 111-2221-E-007-077.

## APPENDIX

In the appendix, we derive the worst-case probability $P$ shown in equation (1) for a successful adversary attack and the approximated probability shown in equation (2).

As discussed in section 4.1, for a successful attack, an adversary must steal the owner's backup copies and try to fool up to $k$ trustees out of a total of $N$ contacts. When an adversary contacts for a request, a person either ignores the request and does not notify the owner or is fooled. The goal of an adversary is to fool $k$ trustees without causing any notification. We assume the adversary randomly selects contacts and asks them one by one. Suppose that the adversary successfully fools the $k$-th trustee after asking $i$ trustee (including the $k$-th fooled trustee), and $j$ regular contacts, the attack success probability $P_{i,j}$ of this scenario is equal to the probability $P_{scenario}$ of encountering this scenario, times the probability $P_{fool}$ that $k$ trustees are fooled without giving notification. We have

$$P_{i,j} = P_{scenario} \times P_{fool}. \quad (6)$$

In detail, $P_{fool}$ is the probability that $j$ regular contacts simply ignore the request (with the probability $p_1$ each), while the $i$-th trustee is fooled. At the same time, $k-1$ trustees among the previous $i-1$ trustees are fooled (with the probability $q_3$ each), while other trustees being requested just ignore the request (with the probability $q_1$ each). We have,

$$P_{fool} = p_1^j \times C_{i-k}^{i-1} \times q_3^k \times q_1^{i-k}, \quad (7)$$

where $C_j^i = \frac{i!}{(i-j)!j!}$ is a binomial coefficient.

$P_{scenario}$ is the number of attack scenarios divided by the number of all possible events. The number of all possible events is $\Omega = C_n^N = \frac{N!}{(N-n)!n!}$. The number of occasions before encountering the $i$-th trustee is $C_j^{i+j-1} = \frac{(i+j-1)!}{(i-1)!j!}$, and that after picking the $i$-th trustee is $C_{n-i}^{N-i-j} = \frac{(N-i-j)!}{(N-n-j)!(n-i)!}$. Therefore, the number of attack scenarios is $C_j^{i+j-1} \times C_{n-i}^{N-i-j}$. We then have

$$P_{i,j} = \frac{1}{C_n^N} \times C_j^{i+j-1} \times C_{n-i}^{N-i-j} \times p_1^j \times C_{i-k}^{i-1} \times q_3^k \times q_1^{i-k} \quad (8)$$

Then the attack success rate we have in equation (1) can be calculated by summing up the probability $P_{i,j}$ of all success scenarios and multiplying the result with the probability $P_{steal}$ of the successful stealing of the owner-possessed secret, i.e.,

$$P = P_{steal} \times \frac{1}{\Omega} \sum_{i=k}^{n} \sum_{j=0}^{N-n} C_j^{i-1+j} \times C_{N-n-j}^{N-i-j} \times p_1^j \times C_{i-k}^{i-1} \times q_3^k \times q_1^{i-k}$$

Since equation (1) is complicated, we proposed an approximation. We observe that for the case of a large $N$, the adversary must request many regular contacts and has a very low probability of having them all ignore the request. Therefore, we may assume that the attacker needs to succeed in the first k tries to locate the trustees who can be fooled. The probability of each contact attempt to find a trustee is equal to $n/N$. With the success rate of fooling a trustee being $q_3$, we have the approximated successful attack rate equal to $(\frac{n}{N} q_3)^k$.

**Wei-Hsin Chang** received a B.S. degree in Electrical Engineering from National Tsing Hua University, in 2014 and an M.S. degree in Electrical Engineering from National Tsing Hua University, in 2018. He is currently working at DeepMentor and taking charge of the hardware platform.

**Ren-Song Tsay**, nicknamed "Dr. Zero-Skew", is the inventor of the famous industry standard zero-skew clock tree design method. He received his Ph. D. degree from UC Berkeley in 1989 and worked for IBM T. J. Watson Research Center before he started his Silicon Valley ventures. He was the person who designed the first commercially successful performance optimization physical design system and then jointly founded Axis Systems and developed a breakthrough logic verification system using reconfigurable computer technology. After that, he helped a few start-up companies as a consultant or investor and is now teaching at National Tsing-Hua University. He received the IEEE Transaction on CAD Best Paper Award in 1993 and IBM Outstanding Technical Achievement Award in 1991.